# Unique support from identity-based groups: Professional networks of women and LGBTQ+ physicists analyzed and compared by career sector

Authors: Chase Hatcher,[1] Adrienne Traxler,[2] Lily Donis,[3] Madison Swirtz,[4] Camila Amaral,[4] Justin Gutzwa,[5] Charles Henderson,[6] Ramón Barthelemy[4]

[1]Integrated Sciences Initiative, North Carolina State University, Raleigh, North Carolina, USA
[2]Department of Science Education, University of Copenhagen, Copenhagen, Denmark
[3]Department of Sociology & Criminology, University of Utah, Salt Lake City, Utah, USA
[4]Department of Physics & Astronomy, University of Utah, Salt Lake City, Utah, USA
[5]Department of Educational Leadership, Michigan State University, East Lansing, Michigan, USA
[6]Department of Physics and Astronomy, North Carolina State University, Raleigh, North Carolina, USA

Abstract: This paper presents a social network analysis of the professional support networks of 100 LGBTQ+ and/or women PhD physicists, comparing the networks based on the career sectors of academia, industry, and government/nonprofit. The methods for constructing and analyzing the ego networks, which are novel in many ways, are explained in greater detail in an earlier publication (Hatcher et. al., 2025). We describe the networks of the whole group and use statistical tests of independence to explore differences between sectors in terms of whole network metrics, network composition based on alter (network member) characteristics, and support. We find that alters associated with groups (like affinity groups and personal and professional interest groups) are more likely to provide identity-based and community building support, participants in Academia have fewer personal friends in their networks while those in Industry have more, participants in Government report less instrumental support, and those in Academia report less material support. These results and others lead to suggestions for employers in these sectors on how to better support these physicists, including continuing to promote participation in affinity and interest groups, providing more material support and/or personal time in the academic sector, and more instrumental support in the form of professional development or training in the government sector.

# I. Introduction

Physics, as a field, has a well-documented history of being un-welcoming to gender and sexual minority scientists (GSM scientists, including anyone who is not a cisgender heterosexual man). Beyond statistics showing underrepresentation of women in varying levels of physics education (more than three-quarters of all bachelors degrees in the US were awarded to men in recent years) [1], studies have shown that GSM physicists experience explicit discrimination and exclusion at all levels of physics education [2,3], which may be a reason why so many choose to leave the field. Such experiences are unfortunately also shared by physicists with other minoritized identities in physics, such as Black physicists and other Physicists of Color [4,5], physicists with disabilities [6,7], and others [8], but the current study focuses on GSM physicists, so our discussion will remain limited to their experiences.

Researchers have found that retention of GSM physicists in physics at least partially comes down to being supported early in one's career and by different kinds of network members (friends and family to peers and mentors) [2,9,10], but most studies that examine support for GSM physicists focus on its absence [8], sometimes calling for more research on how GSM physicists are supported [10]. A better understanding of what support looks like for gender and sexual minority physicists would help us understand what motivates people from marginalized backgrounds in physics to decide to stay anyway. It would also address a gap in the physics community's understanding of its diversity problem by aligning an investigation more with the not-so-common question of, "why do GSM physicists stay?" rather than the more-often investigated question of, "why do GSM physicists leave?" We should be promoting diversity in physics not only on a moral basis of equitable access and inclusion for all, but also because diversity has been shown to lead to better science across the board [11].

This study helps us understand how GSM physicists find professional success by providing a comprehensive understanding of their support networks using tools from social network analysis (SNA). Our team has interviewed 100 women and/or LGBTQ+ physicists about their career trajectories and their support systems, and during the interviews they created diagrams representing their professional support networks. By analyzing the interviews and support networks of these physicists, who span the career sectors of academia, government, and industry, we can get an idea for how their networks function and how support differs based on career sector and other dimensions of personal identity.

Collecting, coding, and organizing visual, qualitative, and quantitative network data from 100 individuals amounted to a complex and somewhat unprecedented task which constrained and steered our analysis in ways we did not anticipate, mostly through non-uniformity in collected data for all participants. In other words, this analysis was largely exploratory and became more targeted as results arose and as avenues for specific investigations were found to be dead-end or promising. Still, this study was guided by a primary research question which we expanded based on specific targets for analysis:

- (RQ1) *How can we characterize the networks of the study participants, and how do these characterizations differ based on participants' career sector?*
    - (RQ1A) *How do we characterize the size, density, and transitivity of the networks, and how do these differ based on career sector?*
    - (RQ1B) *How are the networks composed based on alter attributes, and how does composition differ based on career sector?*
    - (RQ1C) *How do we characterize support, and how does it differ based on career sector and alter attributes?*

With a better understanding of the support systems that GSM physicists have relied on to get to where they are now, we can provide recommendations to employers, professional societies, and institutions on how to better support these scientists going forward.

This study is part of a larger project using the same dataset for different analyses. The research goals originally proposed for the larger project include: (1) developing a "basis set" of sources of professional support to describe physicists'networks, (2) comparing mentoring and career trajectory experiences within GSM identities, and (3) comparing network characteristics between physics workforce sectors. This study primarily addresses the third item and partially addresses the first, while other researchers on this project team intend or intended to address the second.

This paper is meant to accompany and follow an earlier publication, Hatcher et al. (2025) [12] which explains our methods for constructing and analyzing participants' social networks and provides background on the subjects of critical theory and SNA that inform the work. We refer to the earlier publication as the "Methods Paper." The current paper briefly reviews key topics in the Methods Paper and then focuses on results of the analysis, split into general results for the full group and comparative results between participants in the three career sectors of academia, industry, and government. We conclude with a discussion of the results, their implications for the physics community, limitations of the study, and directions for future work. This paper provides a comprehensive understanding of our participants' professional support networks, differences that exist between them based on career sector, and what these results mean for supporting GSM physicists going forward.

# II. Background & Theory

## Careers in Physics

Research on the careers of professional physicists has commonly split their work into the career sectors of academia, industry, and government. These sectors are a refined version of the traditional division between employment in the public ("government," in our terminology, where compensation comes from taxpayers) and private sectors ("industry," where compensation comes from a private institution or individual). Academic institutions, which can be public or private, bridge the gap between these and represent an especially large employment sector for scientists, making it reasonable to treat it as a separate category. Examples of academic positions for physicists would be faculty (tenure-track or otherwise), research, administrative, or other appointments at universities or institutions of higher education. Examples of industry positions include a variety of job titles in engineering, tech, science communication, and consulting. Examples of government positions include those at national laboratories, federal agencies, and non-profits.

In 2024, The American Institute of Physics (AIP), using the same sector divisions, reported that nearly half of all new physics PhDs surveyed from 2021-2022 were initially employed in the academic sector (46%) [13]. About a third (34%) found initial employment in the private sector ("industry," in our terminology), and 16% found initial employment in government while 4% found initial employment in "other" sectors. They also divided positions into "postdoc," "potentially permanent," or "other temporary" and found that most postdoc positions were in academia (72%), most potentially permanent positions were in the private sector (73%), and most other temporary positions were in academia (63%).

AIP, through the work of Porter, also published a more extensive longitudinal study of the careers of physicists with PhDs [14]. The results come from qualitative analysis of two open-ended survey questions related to success and barriers to success for mid-career physicists. They found that physicists in the private sector more often attributed success to skills and abilities (like interpersonal and computing skills), while those in academia and government attributed success to persistence, work ethic, and social support. Likewise, barriers in the private sector also involved skills as well as a lack of career opportunities, while barriers in academia and government were more often problems with the institutions themselves or social problems. This study also compared responses based on gender and found that men more frequently discussed their skills and women more frequently discussed social issues or support in terms of both successes and barriers to success. They also found that men more often discussed organization issues and women more often discussed gender bias in the context of barriers.

Other studies have variously looked at gendered power structures for physicists in the academic sector [15,16], careers of physicists in the private sector [17], training for professional physicists [18], and learning orientation as it relates to productivity for physics PhDs [19]. Many publications concerning careers for PhD physicists are focused on the job market (as bullish [20] or bleak [21]), guidance on career pathways [22,23], or the promotion of a specific field [24]. In other words, while there is research on the career experiences of physics PhDs, much writing on physics careers is directed towards physics students and takes a voice of guidance, whether it is encouragement or deterrence.

This study adds to the physics community's limited knowledge of the experiences of professional physicists in their careers, focusing on those of GSM physicists. This work not only serves as a basis for making suggestions to employers for how to better support these physicists, but can also serve as a guide for aspiring physicists on what professional physics might entail.

## Social Support in Career Studies and PER

Social support is a construct commonly encountered in social network analysis that has been adopted as a theoretical framework for understanding social networks themselves. In a book chapter reviewing social support, Song, Son, and Lin note an entanglement of social support with social cohesion, social integration, and social capital as network theories derived from the network perspective (the use of SNA), and proceed to differentiate them [25]. They identify cohesion as the nature and degree of harmony among network members, integration as the extent of participation in a network, capital as resources embedded in a network, and support as the aid that individuals gain from their network members. They present these theories as a causal chain in the order listed above, with social support "downstream" of all others. In other words, social support (receiving aid) requires social capital (people in a position to offer aid), made accessible through social integration (viable ways to access people offering aid), which all works only with adequate social cohesion (harmonious or at least non-hostile overall affective quality of a network). In an earlier discussion of social support, Lin et al. (1999) assert that it contains two components: structural bases (locations within network structures) and functional elements (specific instances or types of support that may or may not be elicited from the different bases) [26]. Extending from this conception, our application of the framework focuses on thorough descriptions of supportive relationships: we identify the nature of the support (by naming and describing it, attending to *function*) and identify its source (by associating it with a specific alter in a specific professional/personal context, attending to *structure*).

As Traxler et al. (2024, related to this project) point out, a theoretical framework in the context of SNA hopefully answers the question, "Why do you think the network is important?" [27] In this study, we think the most important thing about the networks of participants is the social support they provide towards a successful career in physics. That is not to say that other concepts like social cohesion, integration, capital and their associated theories are not important, but since social support can be seen as descendant from and reliant on these concepts, we see it as the most succinct and holistic way to understand the networks. Indeed, the original project proposal considered both social support and social capital as viable theoretical frameworks, but through the data collection process, the focus switched primarily to support as a more practical framing for the interview protocol and as a more useful vocabulary in the coding process. We found it easier and more informative to ask about and code for the presence or absence of different types of support rather than the strength of ties, which is more common in social capital studies. In doing so, we named 9 types of support as they appeared in the interviews since there is little consensus in the literature on a basis set of social support types beyond emotional and instrumental support [25,28] (we define these in Section IV). Furthermore, social support is uniquely relevant to the study of career advancement for marginalized professionals and of success for marginalized physicists, as we explore in the remainder of this subsection.

As Song et al. point out, social support has an extensive history of application in health sciences, namely epidemiology, psychiatry, and psychology, but less so in sociology, education, and career studies [25]. Still, some researchers have reviewed its applications in education contexts, such as Mishra (2020), who focused on underrepresented students and made the case for social support as a particularly viable lens with which to view their experiences [29]. Mishra's review of 136 studies reveals that social support is a critical factor in minority students' academic success in higher education, where different types of support received from different types of network members (i.e., advice and encouragement from parents, emotional support from peers with a similar background) complement each other in a necessary but often still lacking way (compared to other students). Although this work focuses on students rather than professionals, it points out that unique social supports arise from the different paths minoritized students take through higher education compared to mainstream students. We find a similar trend in social support research in career studies, where the framework appears uniquely apt for understanding the experiences of professionals overcoming barriers in their careers.

In a review of studies focusing on social support in the workplace, Jolly et al. characterize social support as a "valuable resource that drives behavior, influences affect, and can provide a buffer against stressful job demands." [28] Though their review focuses on integrating various definitions and theoretical lenses in the studies they review, their characterizations of the function of social support across studies capture the value of social support as a lens for a study like this, particularly in helping professionals overcome challenges they encounter at work. Indeed, other studies have focused specifically on the value of social support in reducing workplace burnout, highlighting the importance of emotional support [30] and support from families [31]. Similarly, social support has been used to explain career success [32], satisfaction [33], and optimism [34] for minority professionals, and in all cited cases was closely linked to intentional professional networks like the affinity and identity-based groups we discuss further in this paper. Social support has seen various applications in career studies, and while not always from a networks perspective (sometimes from a psychological/mental health perspective), many applications (especially those focusing on mental health) demonstrate the value of this theoretical lens in understanding professionals' perseverance through adversity, which is why we think it is suited to our study of GSM physicists.

Our adoption of social support as a framework means that we are using support as a lens through which we understand the structure and function of participants' networks. This is reflected in our approach to constructing and analyzing the networks, where support was a major theme in both the interview protocol and thematic coding process. Our focus on support is guided by PER literature which has identified social support as a common thread in factors leading to greater retention and success for underrepresented physicists. Studies have highlighted support from faculty and other senior researchers [10,35] family/parental/spousal support [9,15], support from affinity or identity-based groups [4,5], and support from outside the classroom/academic institution [5,36] as critical for women physicists' success at various levels. More commonly, however, studies that examine support for women and other underrepresented physicists focus on its *absence* [5,8,15,37], which may leave us wondering what that support actually looks like (function) and where it comes from (structure).

The state of research on support for GSM physicists as described above is exactly why a study like this is important—we want to provide a more detailed description of support for GSM physicists from a networks perspective, relying on social support as a framework. Doing so fills a gap in the literature where research on support has (mostly) focused on its absence or has discussed it without much specificity. That is not to say this research is not helpful or important, only that it does not paint a vivid picture of support. By using a social support framework, we view both the structure and function of these physicists' networks through a lens of support and therefore see their "success" (the fact that they have a PhD in and are working in physics or a related field) as at least partially dependent on the support they received from their network. We do not assert that all members of participants' networks served to support them (see Gutzwa et al., 2024, for contrasting examples, especially in graduate education [2]) or even that the support they received always translated to professional success, but based on theories of social support and on research around what support does for GSM physicists, we suppose that all participants' networks did, in some way, support them towards a career in physics, and we draw conclusions from and build interpretations of results accordingly.

## Critical Theory

The larger project, of which this study is a part, is concerned with cultural and identity-based experiences in physics spaces and was conceived in light of a documented history of physics spaces being discriminatory and exclusive. To that end, we have used critical theory as a basis from which to investigate questions related to the experiences of people with marginalized identities in physics. Critical theory is a mode of inquiry and body of sociocultural theory that has historically been oriented towards the betterment of life for oppressed groups [38]. Today, sub-research areas such as feminist studies, queer and gender studies, critical race theory, and ethnic studies are considered by some to exist under the umbrella of critical studies, even though these areas have distinct traditions and practices that separate them.

Underlying many works of critical theory is the rigorous examination of power and the way it functions in society to control behavior by normalizing some actions and attitudes and demonizing others [39]. Critical theorists have consistently demonstrated the socially-constructed nature of social norms, showing that deeply rooted sociocultural value systems (such as race [40], patriarchy [41], and heteronormativity [42]) are constructed and maintained by the ruling classes to serve their interests rather than being universal or natural. Critical theories invite us to challenge widely held assumptions and beliefs to discover their origins and who they may serve, often leading to liberatory and revolutionary ideas for oppressed peoples.

Critical theories also shape research practices and methodologies. In the case of this study, the approach to network construction, network analysis, and reporting and interpretation of results are all designed to be aligned with guidance from critical theorists. The Methods Paper provides more background on critical theories relevant to this project and study (including critical work in PER) and the way that critical considerations have factored into the methodology. An earlier publication from this project, Gutzwa et al. describe the role of critical theories in shaping the theoretical foundation of this project overall [2]. In short, our network construction procedures are based on a maximal amount of input from participants so that the networks we construct center the participant and their perspective and are as faithful as possible to those that actually exist for participants. We rely heavily on qualitative data sources so that the words of our participants are the primary means by which their experiences are depicted, something which is especially important for understanding experiences with discrimination and marginalization. Finally, our reporting process includes attention to the way that power is constructed and maintained socially in physics contexts and is guided by the overarching goal of improving the experiences of marginalized physicists.

## Social Network Analysis

Social network analysis, or SNA, is a set of tools and techniques for analyzing groups of people based on their connections with one another, qualities of the people, and qualities of the connections themselves. It typically involves the creation of network visualizations (sociograms) with people represented by shapes like dots (called nodes) and lines between them representing connections (links or edges). It also usually involves the calculation of various metrics that describe characteristics of individuals or of the whole group. SNA can involve a mix of quantitative and qualitative techniques in constructing and then analyzing the networks and comes in a few modalities which are worth reviewing here.

Typically, and especially in PER, quantitative approaches to SNA are more common. As described in Traxler et al. 2024, quantitative approaches refer to a reliance on quantitative methods in the analysis of networks [27]. This often involves the calculation and comparison of various metrics that describe characteristics of networks. Alternatively, researchers could take a more qualitative approach to analysis, choosing instead to analyze networks based on descriptions from members, or they could take a mixed-methods approach, mixing quantitative and qualitative methods. Some researchers have promoted mixed-methods SNA (MMSNA) as a best of both worlds approach that combines the analytical power and scalability of quantitative methods with qualitative methods' ability to capture nuance and provide mechanistic explanations for the networks [43].

SNA studies in PER (and more broadly) also typically take a sociocentric approach to network construction, meaning that they aim to represent a network that exists among a bounded group of people, usually based on all members' descriptions of their connections. Alternatively, we may take an egocentric approach to network construction, whereby one person (the ego) describes their relationships with others (alters), forming a network based around one person. That network is treated as independent of any others, meaning that a collection of ego networks lends itself well to inferential statistics. There are other advantages and disadvantages to each of these approaches, so their use would mostly depend on project goals and research questions.

The Methods Paper provides more background on SNA and how it has been used in PER and it includes a lengthier discussion of decisions made concerning available SNA modalities. In short, we chose to use MMSNA and egocentric SNA approaches because together they: (1) are more appropriate for our research questions which are based on our study participants' experiences in physics, (2) center our study participants in all aspects of their network, including its construction, and (3) allow for the exploration of complex sociocultural phenomena through the words and perspectives of our participants.

# III. Methods

Methods for constructing, analyzing, and visualizing participants' networks are described in detail in an earlier publication [44]. We summarize those methods here and expand on features especially relevant to this analysis.

## Recruitment, Demographics, and Non-comparative Considerations

The goal of recruitment was to compile a sample of PhD-holding physicists who work in the US in a physics or physics-related field, with roughly equal representation of Women and LGBTQ+ physicists and physicists in academia, industry, and government. Participants self-reported personal identities in a demographics survey that included select any/all options. Recruitment started with the researchers' personal networks and suggestions from the project advisory board (representing all three sectors) and quickly graduated to calls for participants in relevant email listservs, professional society websites, and on social media. Snowball sampling and word of mouth also contributed to the participant pool, and as the sample grew, researchers were selective about additional participants to admit in order to maintain balance between gender/sexual identity and career sector (though it was hard to avoid over-representation of the academic sector). As such, this is not a random sample of GSM physicists but rather a sample which aims to roughly equally represent the qualities and identities we were interested in comparing and describing. Figure 1 illustrates the participant demographics.

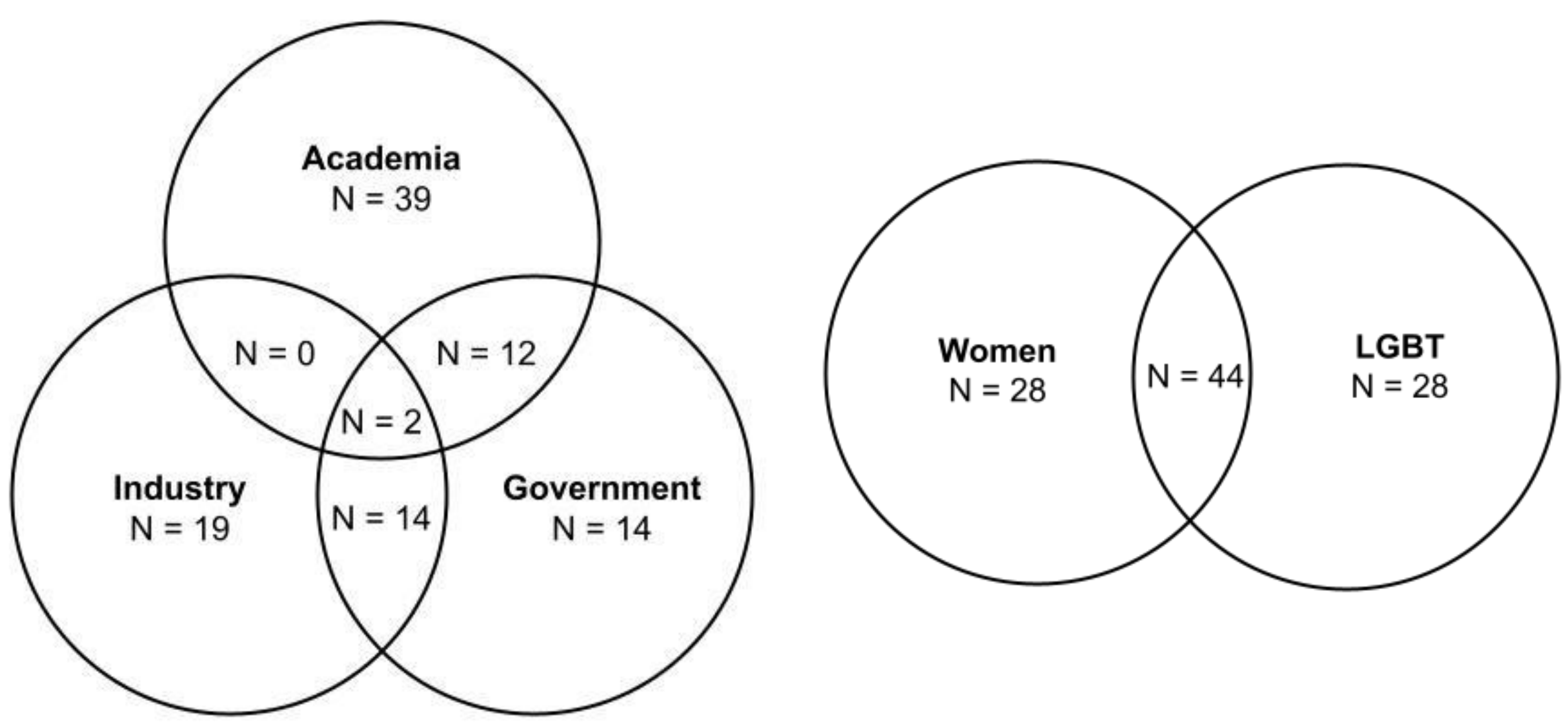

Figure 1: Venn diagrams representing career sectors (left) and gender/sexual identities (right) of participants.

As described in more detail in the Methods Paper, this study was conceived in light of an apparent trend in PER of taking a comparative approach to identity-based research, whereby a minority group is consistently compared to the majority group, usually revealing some lack, deficit, or structural barrier. This study aims to subvert this trend by only examining the minority group and only making comparisons within that group, thereby developing an understanding of GSM physicists' experiences based only on their experiences rather than the experiences of the majority group.

Without comparing to a larger (majority) group of physicists, we are not able to say whether the qualities we observe in this sample are different from the rest of the population of physicists or not—we can only describe this sample and make claims concerning differences found within the sample. This approach limits comparative claims, but we see it as a first step in the field's development of a basis for understanding the support networks and professional experiences of GSM physicists. Future work may build on this foundation by collecting similar data from majority groups for comparison.

## Network Construction, Analysis, and Visualization

The networks originate from interviews conducted with 100 women and/or LGBTQ+ PhD-holding physicists. The interviews served as both a source of qualitative data concerning our participants' professional experiences as well as a name generator (network construction tool) for their support networks. The interviews contained questions related to participants' career journeys and professional trajectories and the ways they were supported (or not, "gaps in support") along the way. They included a sociogram construction component, wherein participants would add names to sticky notes on a Google Jamboard and place them on a set of concentric circles with a star in the center representing the interviewee (a sample sociogram is shown in Figure 2). These sociograms, combined with thematic coding of interview content relating to alters placed on the sociogram, comprised the data that went into our network construction.

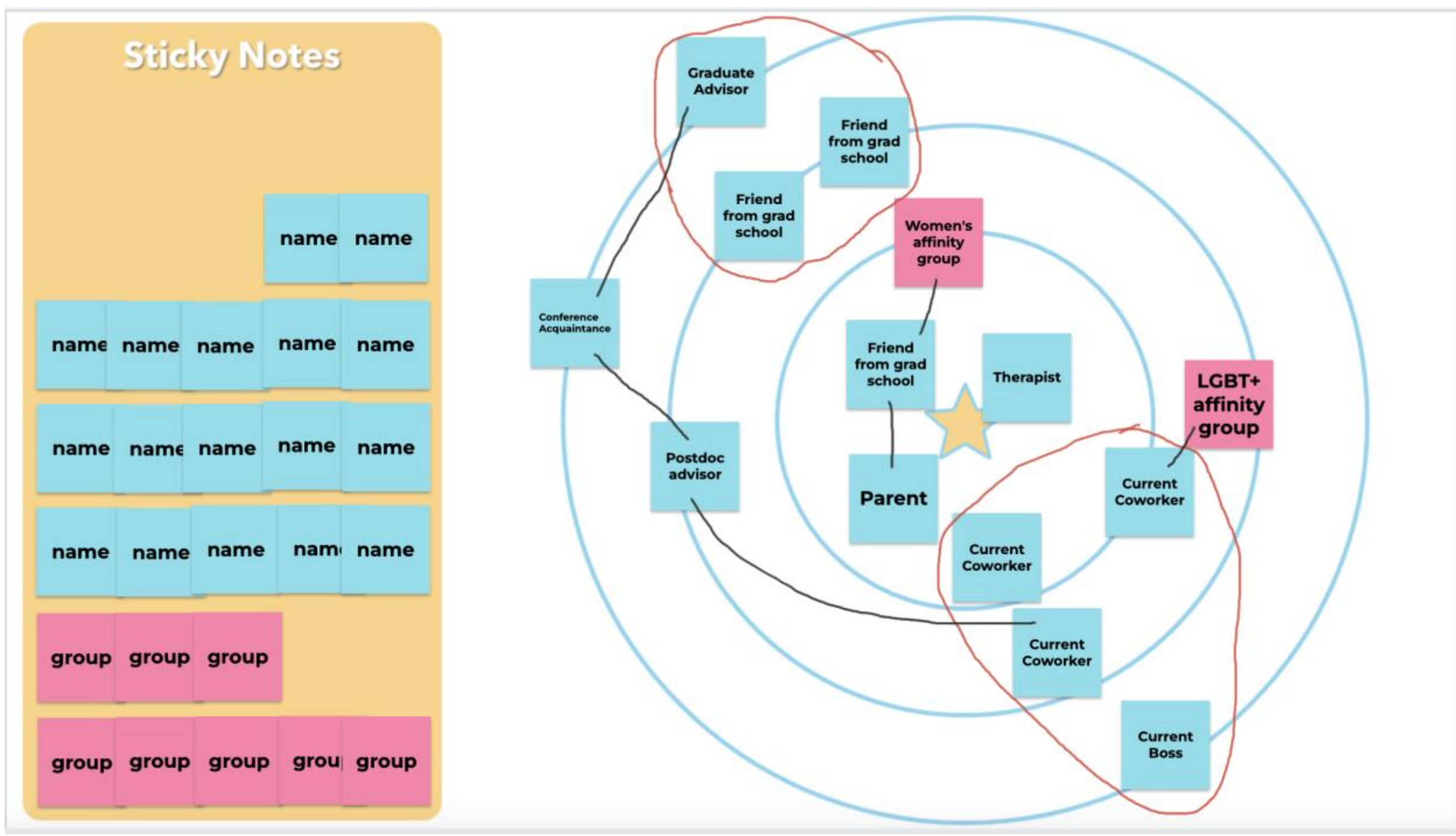

Figure 2: Fictionalized example of a sociogram showing alter connections. Individuals were added to blue notes while groups, organizations, or institutions still treated as single nodes were added to pink notes.

After an initial round of thematic coding, which included coding for network data such as alter relationships and associated types of support (shown in Table 1), the interviews underwent a second round of network-oriented coding that further specified relationship and support codes for every mention of an alter. All coding took place after the interviews and without direct input from interviewees. Ultimately, the various coding processes led to a list of every alter placed on a sociogram specifying all professional and personal relationships that person had with the ego, all types of support (or gaps in support, a noted absence of support) they were associated with, the number of times they were mentioned in the interview ("Mentions"), their proximity to the center of the sociogram ("Proximity"), and whether they were ever associated with "Dissonance" or some sort of interpersonal tension or friction with the interviewee. We also created a list of "who to whom" connections among alters (and including connections between the ego and all alters) based on connections among alters indicated by participants on the sociograms.

Table 1: Network data codes, a subset of the full codebook available as Table 1 in the Methods Paper. Note abbreviations for support codes which feature heavily in the rest of the paper.

| 1st level | 2nd level | 3rd level |
|---|---|---|
| Connection Nodes | Individual | Partners<br>Friends<br>Family<br>Other |

| | Group | Subfield<br>Group/listserv<br>Conferences<br>Affinity group |
|---|---|---|
| | Institutional | Undergraduate program<br>Graduate program<br>Postdoc<br>Past job<br>Current job<br>Future job |
| | Collegial & Supervisorial | Bosses/supervisors<br>Mentors (professional)<br>Mentees/direct reports<br>Coworkers/peers |
| Nodal Support | Identity-based support (ID)<br>Physical closeness (Phys. Cl.)<br>Networking support (Netw.)<br>Community building support (Comm.)<br>Career advice support (Car. adv.)<br>Emotional support (Emo.)<br>Material support (Mat.)<br>Instrumental support (Inst.)<br>DEI initiatives/policies (DEI)<br>Gaps in support | |

With a list of ego attributes from a pre-survey (including job title, career sector, gender identity, and more), alter attributes from thematic coding for network data described above, and connections between any two nodes, we were able to construct networks in R using the igraph and egor packages.[34–36] Some network visualizations included a maximal amount of information, including support encoded in edge color and thickness, while some simplified networks into uniform shapes or areas to make comparison easier.

In addition to visualizing each network in various ways, we were also interested in comparing networks based on descriptive metrics. The full list of metrics is described in Table 2 of the Methods paper. From this list, we selected metrics to focus our analysis on based on our research questions. The final selection focused on basic parameters of interest in SNA which we also expected to be contextually informative. For example, density is a common measure of global connectivity in networks, and in this study it would tell us whether participants' alters tended towards isolation or embeddedness with one another. Transitivity would tell us whether connected alters tended towards tight clusters or not. Study-specific metrics were also developed based on their potential informativity around support: Support to Alters ratio and Gaps in Support to Alters ratio both tell us how supportive (or un-supportive) participants' networks may be. In general, the metrics we aim to describe and/or compare were chosen based on some combination of their ubiquity, ability to translate to meaningful characterizations of networks, ability to speak to our study goals, and where we expected to see differences among career sectors based on the interview data.

Table 2: Selected characteristics and metrics describing whole networks and their associated definitions. All are calculated without the ego included in the network[48].

| **Network characteristic or metric** | **Description** |
|---|---|
| Total Nodes | Number of alters |
| Density | Global density, or total number of connections divided by total number of possible connections (igraph's edge_density function) |
| Transitivity | Global transitivity (sometimes called clustering coefficient), or the probability that adjacent nodes are connected (igraph's transitivity function) |
| Support to Alters Ratio | Ratio of Total Support, or number of instances of support (each alter could have up to 9 associated types of support) to Total Nodes |
| Gaps in Support to Alters Ratio | Ratio of Total Gaps in Support (similar to Total Support) to Total Nodes |

## Alter Classification

In order to compare participants' networks, it was helpful to classify alters into categories. In general, we were interested in any alter attributes and demographics that could tell us something about their role or place in ego's network, especially if those qualities were professionally relevant, but we were limited by the consistency of data across interviews (for example, we knew the age of only some alters, meaning age was not a viable attribute for comparing alters and networks). The following classification schemes were developed because of their relevance in professional and academic social dynamics and because they could be attributed to all alters across all interviews.  Some of this approach was discussed in the Methods Paper, but we review it here in greater detail for this analysis.

Alters were classified based on the career stage of their associated ego in which they were first mentioned ("Stage") and the primary nature of their relationship ("Relationship") with their associated ego. Classification based on Stage was relatively straightforward as we had already coded each mention of an alter with their affiliation where appropriate (Table 1, Connection Nodes > Institutional). Since undergraduate, graduate, post-doctoral, past job, current job, and future job affiliations generally follow a linear and progressive path through someone's educational and career stages including a PhD program, we classified each alter's Stage based on the earliest role with which they were associated. The final Stage categories were "undergrad," "grad," "postdoc," "job," and "none." We combined the three phases of job codes because the time one might spend in any of these positions would vary more than the academic positions, and since they would all comprise a professional career stage. The "none" category included any alter with no affiliation codes, which was common for friends, family, and others with relationships to the ego which were not based primarily in their professional/academic career. Stage ends up roughly representing the age of an alter's relationship with ego, measured by academic and professional benchmarks.

Classification based on relationship type was slightly more complicated, since the Individual, Group, and Collegial & Supervisorial sub-codes (Table 1) were not mutually exclusive and there were too many to allow for straightforward comparison. Instead of using these directly, we created 8 classes which we assigned to alters based on the logic described in Table 3. The inclusion criteria were determined during meetings of most of the authors and were meant to represent somewhat exclusive and common categories of relationships we observed in the interviews. We note that they are post-hoc classifications that are based on the words and indications of our participants but are assigned according to logic determined by the researchers.

Table 3: Relationship categories (some with abbreviations) and their inclusion criteria.

| **Relationship category** | **Inclusion criteria** |
|---|---|
| Peer | This is the default. Any alter that does not meet criteria for another category is defined as this. |
| Mentor | Coded as Mentors (Professional) and does not meet criteria for any following categories in this table. |
| Groups | Coded as Group/listserv or Affinity Group and does not meet criteria for any following categories in this table. |
| Friend - professional (Friend.prf) | Coded as Friends and does not meet criteria for any following categories in this table. |
| Friend | Coded as Friends and NOT coded as Subfield, Conferences, or Coworkers/peers, and does not meet criteria for any following categories in this table. |
| Family/partners (Fam.part) | Coded as Family or Partners and does not meet criteria for any following categories in this table. |
| Mentees | Coded as Mentees/direct reports and does not meet criteria for any following categories in this table. |
| Boss | Coded as Boss/supervisor. |

## Network Comparison

As described in Section I, the larger project is interested in comparing participants' networks according to several dimensions of personal and professional identity, while this paper focuses on comparison on the basis of career sector. We split analyses across multiple papers because: (1) the different theoretical backgrounds used to interpret results of the different comparisons would make for an exceedingly long paper if combined and (2) we divided up work on the project team so that analyses undertaken by individual graduate student researchers had a realistic scope. In order to compare participants' networks in a comprehensive and rigorous way, we introduced and developed additional methods, which we describe here.

Violin plots provide a way to visualize and compare multiple distributions of a variable for different populations [49]. Like a boxplot, which visualizes distributions by focusing on the relative location of means and quartiles within a dataset, a violin plot can show mean and quartile positions explicitly and/or implicitly by adding curves to reflect relative number of counts at each position. The result is a symmetric and curved shape that might resemble a violin body, especially for bi-modal distributions.Violin plots can help us compare distributions of whole-network metrics across different factors or categories of networks, like those belonging to participants in different career sectors. They can also compare distributions of more specific metrics, like counts of different types of alters for different classes of egos. In Section IV, we use violin plots to visually compare distributions of significantly different parameters.

ANOVA and t-tests are both widely used statistical tests of difference. ANOVA tests assume independence of observations, normality in the distributions of residuals for small ($n \leq 30$) samples, and homoscedasticity or equivalent variances among the groups [50]. Student's t-test (used for comparing just two groups) carries the same assumptions, except normality is expected for the samples themselves and not just their residuals. Both tests also expect to compare continuous quantitative data (the samples) based on discrete categorical factors (career sectors of egos, for example).

Chi-squared tests are used to test for independence between two or more variables based on the relative frequency or counts of different categorical factors composing one of them to a categorical separation of groups in the other. Chi-squared tests require mutually exclusive categories in both variables, meaning that any observation could only be placed in one combination of categories. They also assume independence of the categories, as in ANOVA and t-tests, and are valid as long as at least 80% of cells in the contingency table have at least 5 counts [51].

Results of all of these tests are determined by examining p-values, which tell us the likelihood of the occurrence of the observed data if the null hypothesis is true. Therefore, a sufficiently low (usually $< 0.05$) value suggests that we should reject the null hypothesis, that there is no relationship between the variables tested. When running multiple tests of independence, especially in an exploratory manner, it is important to account for multiple testing effects, where a low p-value is likely to appear simply due to the large number of tests being run [52]. A common and conservative approach to such corrections is the Bonferroni method, where either the threshold for significance is reduced by dividing by the number of tests run in that instance or the p-values are increased by multiplying by the number of tests (the result is the same). We report Bonferroni-adjusted p-values whenever multiple tests are used, meaning that we multiply the raw p-value by the number of tests run in that instance. In our case, multiple testing effects are most relevant when performing post-hoc tests between every pair in comparison groups, sometimes as many as 72 at once. Keeping with common practices of reporting p-values with asterisks to denote their significance, we will adopt and extend the widely used American Psychological Association (APA) code to: * for $p < 0.05$, ** for $p < 0.01$, *** for $p < 0.001$, **** for $p < 0.0001$.

# IV. Results

We begin by reporting results for the full group, focusing on network metrics, alter composition, and support, then we move to a comparison of most of the same metrics on the basis of career sector.

## Full Group Results

### Average Whole-network Metrics

Table 4 shows overall mean values and standard deviations for the chosen network characteristics and metrics, addressing RQ1A.

Table 4: Mean values and standard deviations for all network metrics for the full group.

| Network characteristic or metric | Mean | Standard Deviation |
|---|---|---|
| Total Nodes | 16.8 | 7.3 |
| Density | 0.196 | 0.13 |
| Transitivity | 0.677 | 0.28 |
| Support to Alters Ratio | 1.56 | 0.55 |
| Gaps in Support to Alters Ratio | 0.109 | 0.17 |

### Summary of Network Composition

All networks were composed of alters, each with a categorical description of their “Stage,” or the earliest career stage of the ego with which they are associated, and their “Relationship,” or a general description of the relationship they have with ego. Figure 3 shows network composition in terms of Stage and Figure 4 in terms of Relationship, addressing RQ1B. The pie charts show fractions of the 1680 total alters.

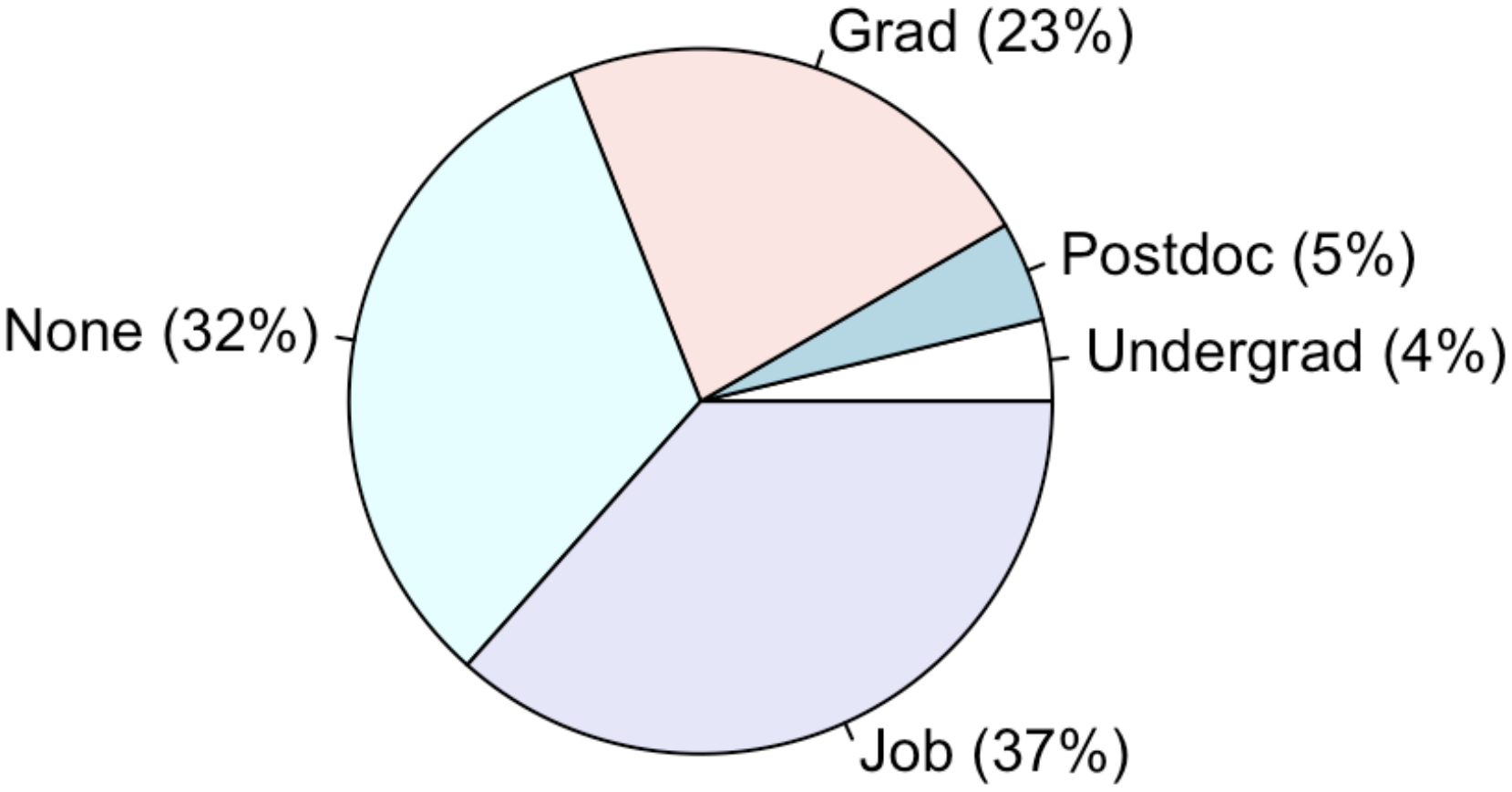

Figure 3: Network composition based on Stage for all participants. Pie slices are arranged by size.

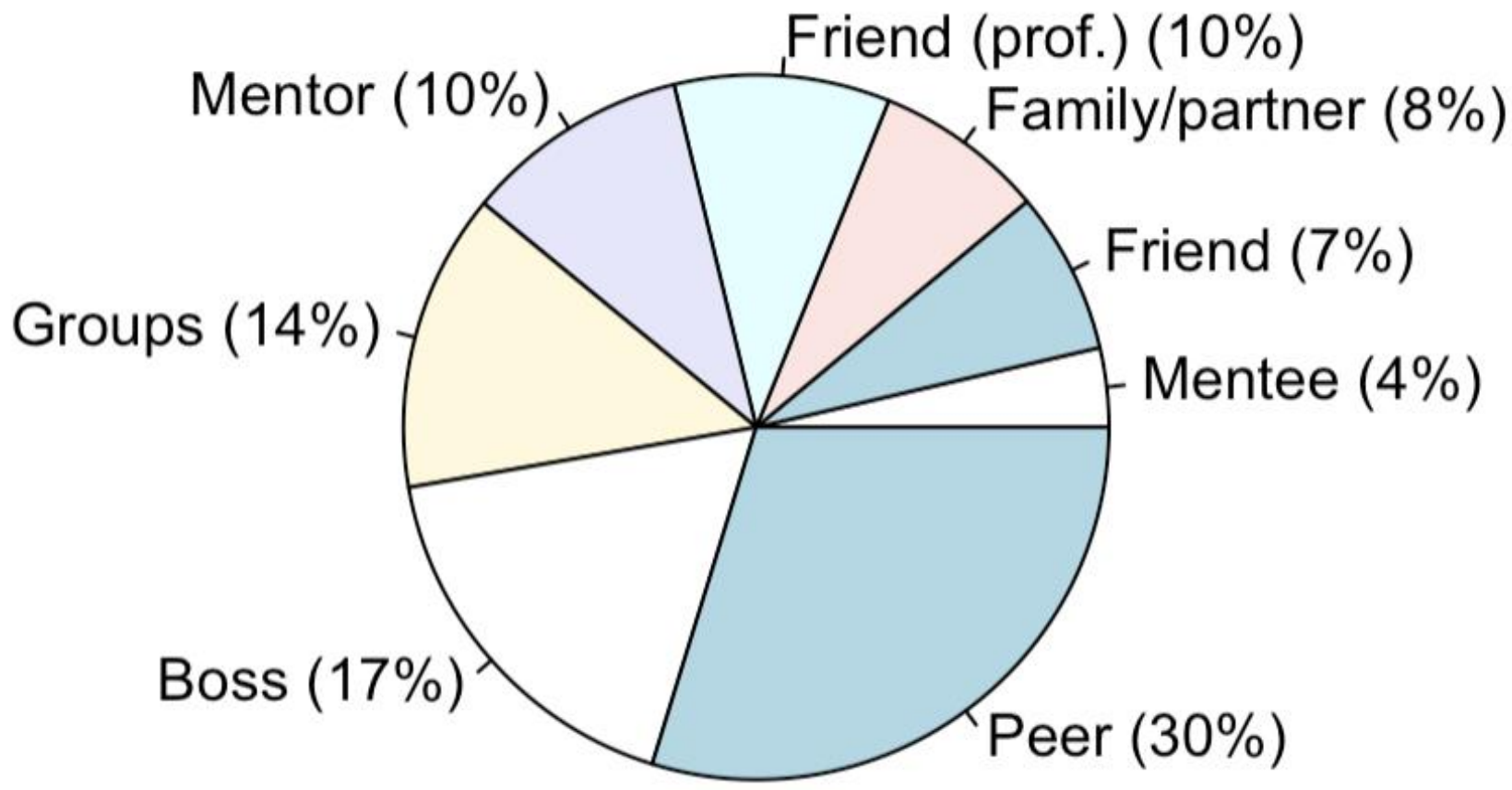

Figure 4: Network composition based on Relationship for all participants. Pie slices are arranged by size.

## Summary of Support Types

Each alter was associated with anywhere from 0 - 9 of the 9 types of support for which we coded. Most of these support types are clearly described in the name, but to be more clear on a few: instrumental support is support with routine operational tasks, like using a copier, learning a new analysis technique, or getting coverage for a class; identity-based support is affirmation, empathy, or camaraderie specifically related to one's marginalized identity; and material support is usually related to financial resources or employment benefits. We examined the relative prevalence of each type of support using pie charts of counts of alters associated with each type across the full group (addressing RQ1C), shown in Figure 5. The total number of support codes was 2490.

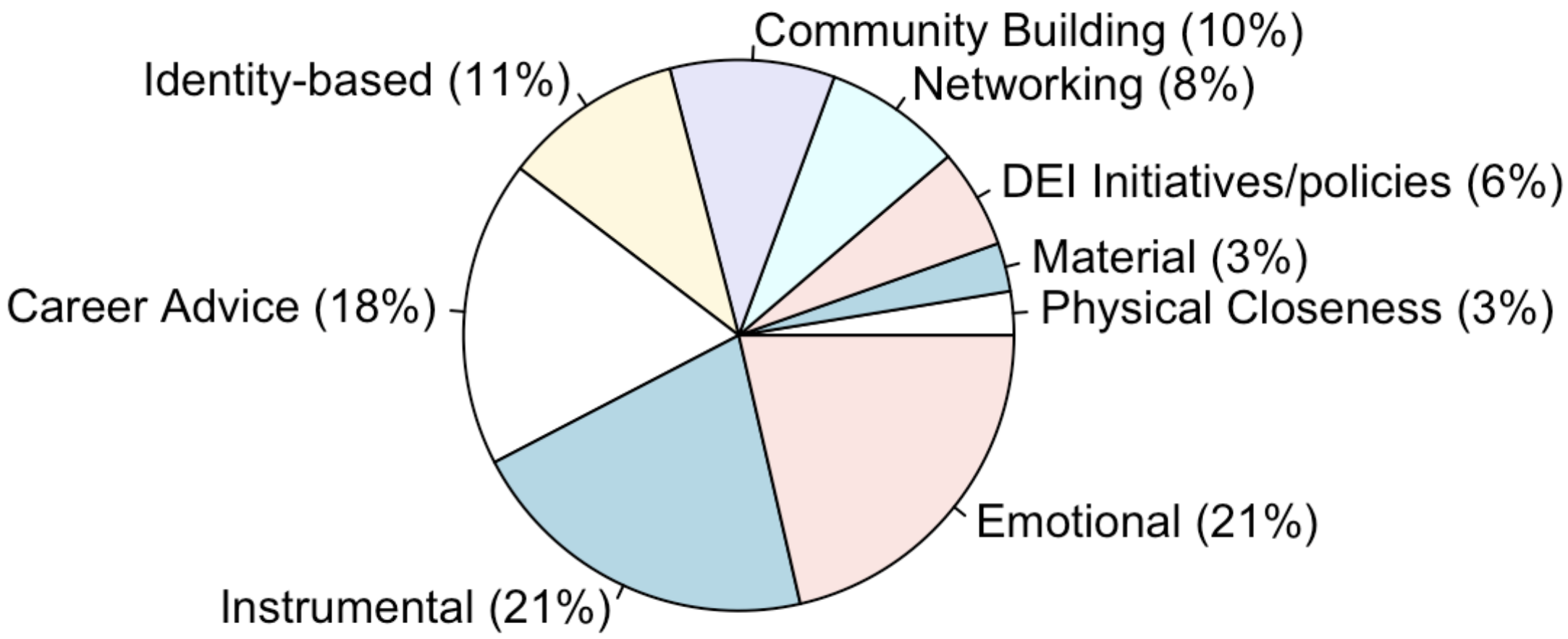

Figure 5: Pie chart showing counts of different support types for the full group.

## Support by Alter Type

We tested if alter Stage or Relationship were associated more or less often with types of support, addressing RQ1C. Tables 5 and 6 show percentages of alters of different classes in the Stage and Relationship categories providing different types of support. While some cells have few observations, both tables still meet the chi-squared test requirement of at least 80% of cells having at least 5 observations [51].

Table 5: Percentages of alters of different Stage classifications associated with different types of support. Percentages give the alter composition of a single type of support, so rows add to 100 while columns do not. Abbreviations for support are defined in Table 1.

| **Support type (N)** | **Undergrad (63)** | **Grad (381)** | **Postdoc (77)** | **Job (614)** | **None (545)** |
|---|---|---|---|---|---|
| ID (264) | 5.7 | 25.4 | 3 | 37.9 | 28 |
| Phys. Cl. (64) | 9.4 | 14.1 | 10.9 | 45.3 | 20.3 |
| Netw. (202) | 4 | 41.1 | 7.4 | 27.2 | 20.3 |
| Comm. (241) | 5.8 | 22.4 | 6.6 | 36.5 | 28.6 |
| Car. adv. (448) | 6 | 25.7 | 7.8 | 33.9 | 26.6 |
| Emo. (533) | 5.6 | 26.5 | 3.9 | 29.6 | 34.3 |
| Mat. (71) | 4.2 | 23.9 | 7 | 47.9 | 16.9 |
| Inst. (522) | 3.6 | 19.7 | 5.7 | 55.6 | 15.3 |
| DEI (145) | 2.1 | 20 | 5.5 | 48.3 | 24.1 |

Table 6: Percentages of alters of different Relationship classifications associated with different types of support. Percentages give the alter composition of a single type of support, so rows add to 100 while columns do not. Abbreviations for support are defined in Table 1.

| **Support type (N)** | **Boss (294)** | **Fam.part (131)** | **Friend (125)** | **Friend.prf (166)** | **Groups (230)** | **Mentee (61)** | **Mentor (173)** | **Peer (500)** |
|---|---|---|---|---|---|---|---|---|
| ID (264) | 13.6 | 4.9 | 8.7 | 8.3 | 31.8 | 6.4 | 10.2 | 15.9 |
| Phys. Cl. (64) | 12.5 | 6.3 | 9.4 | 29.7 | 6.3 | 10.9 | 6.3 | 18.8 |
| Netw. (202) | 28.7 | 2.5 | 5 | 10.9 | 16.3 | 1 | 14.4 | 21.3 |
| Comm. (241) | 4.1 | 5 | 11.2 | 15.4 | 38.6 | 3.3 | 7.1 | 15.4 |
| Car. adv. (448) | 29.9 | 8.3 | 5.8 | 11.8 | 7.1 | 1.3 | 15.2 | 20.5 |
| Emo. (533) | 8.6 | 16.5 | 16.1 | 20.1 | 9.4 | 3.2 | 10.3 | 15.8 |
| Mat. (71) | 31 | 21.1 | 8.5 | 7 | 8.5 | 5.6 | 8.5 | 9.9 |
| Inst. (522) | 28.9 | 3.3 | 4.4 | 12.1 | 5.7 | 3.1 | 15.1 | 27.4 |
| DEI (145) | 13.1 | 0 | 2.8 | 6.2 | 47.6 | 7.6 | 3.4 | 19.3 |

Chi-squared tests of independence of the alter classes within each of these two groups resulted in adjusted p-values $< 0.0001$ for both tests. These suggest that we should reject the null hypotheses, which are that alters classed by their Stage and Relationship, respectively, are not different based on the types of support with which they are associated. These results called for a post-hoc analysis of both groups to determine exactly which combinations of classes of alters and types of support suggest differences within the groups. The 72 post-hoc tests revealed the following significant differences in the Stage category, with asterisks indicating significance levels of each result:

- Grad alters were associated with more networking support.****
- Job alters were associated with more instrumental support**** and less emotional**** and networking* support.
- None alters were associated with more emotional support**** and less instrumental support.****

For the Relationship category:

- Boss alters were associated with more instrumental,**** career,**** and networking* support and less community building**** and emotional**** support.
- Family/partner alters were associated with more material** and emotional**** support and less instrumental** and DEI-related* support.
- Friend alters were associated with more emotional support**** and less instrumental support.*
- Professional friend alters were associated with more physical closeness** and more emotional**** support.
- Group alters were associated with more identity-based**** and community building**** support and DEI initiatives/policies**** and less career advice,**** emotional,*** and instrumental**** support.
- Peer alters were associated with more instrumental support.****

## Sector Comparison Results

A primary goal of this project is concerned with differences in participants' experiences that are based on their career sector. In this section, we show results of comparisons of network metrics, network composition, and support types by career sector.

### Sector Classification

Egos are classified into the three career sectors of Academia, Industry, and Government (capitalization indicating classifications in our dataset rather than the actual sectors) based on self-reported true/false answers to three survey questions about their working in each of the sectors. This means that egos could be affiliated with any of 7 mutually exclusive combinations of 1 - 3 of the sectors. Counts of egos classified in each mutually exclusive sector (combination) are shown in Table 7.

Table 7: Number of egos affiliated with each mutually exclusive sector (combination).

| Sector (combination) | Number of egos |
|---|---|
| Academia | 39 |
| Industry | 19 |
| Government | 14 |
| Academia, Industry | 0 |
| Academia, Government | 12 |
| Industry, Government | 14 |
| Academia, Industry, Government | 2 |

We note the small number of egos affiliated with some of the combined sectors, which makes statistical tests of comparison less informative. In order to work with sufficiently large groups that our statistical comparisons are meaningful, and because we found no significant differences in network metrics and characteristics based on these sector combinations, we choose instead to compare on the basis of affiliation with each of the three sectors rather than the specific sector (combination) indicated. This means that each comparison we describe is based on affiliation with one sector, since the IN and OUT groups for each sector comprise the full group, just split in different ways. Note: IN and OUT refer only to group affiliation, not to direction of ties and edges as the words are sometimes used in network analysis. Counts of egos in or out of each sector, regardless of their affiliation with any other sector(s), are shown in Table 8.

Table 8: Number of egos affiliated with (IN) or not affiliated with (OUT) each sector.

| Sector and counter-sector | Number of egos |
|---|---|
| In Academia ($A_{IN}$) / Out of Academia ($A_{OUT}$) | 63 / 37 |
| In Industry ($I_{IN}$) / Out of Industry ($I_{OUT}$) | 35 / 65 |
| In Government ($G_{IN}$) / Out of Government ($G_{OUT}$) | 32 / 68 |

## Network Metrics Comparison

Mean values and standard deviations of each metric for each group based on sector affiliation are shown in Table 9, addressing RQ1A.

Table 9: Mean values and standard deviations (in parentheses) of selected metrics based on sector affiliation.

| **Sector** | **Total Nodes** | **Density** | **Transitivity** | **Support to Alters Ratio** | **Gaps in Support to Alters Ratio** |
|---|---|---|---|---|---|
| $A_{IN}$ | 17.0 (7.8) | 0.200 (0.13) | 0.660 (0.28) | 1.58 (0.58) | 0.092 (0.12) |
| $A_{OUT}$ | 16.4 (6.4) | 0.183 (0.12) | 0.707 (0.28) | 1.52 (0.51) | 0.137 (0.23) |
| $I_{IN}$ | 16.1 (7.0) | 0.213 (0.15) | 0.684 (0.28) | 1.56 (0.55) | 0.117 (0.23) |
| $I_{OUT}$ | 17.2 (7.5) | 0.183 (0.11) | 0.674 (0.28) | 1.56 (0.55) | 0.104 (0.12) |
| $G_{IN}$ | 16.7 (6.1) | 0.198 (0.12) | 0.686 (0.27) | 1.57 (0.65) | 0.144 (0.23) |
| $G_{OUT}$ | 16.8 (7.8) | 0.191 (0.13) | 0.673 (0.28) | 1.56 (0.50) | 0.092 (0.13) |

To test for differences between groups in or out of each sector, we used three separate T-tests per metric since we only wished to compare two groups each time: a sector's IN and OUT groups. The results of the T-tests (specifically the p-values) tell us whether we should reject the null hypotheses, that there are no differences between the IN/OUT group for each metric tested. None of these tests resulted in P-values smaller than the common threshold of 0.05, meaning that we should not reject any of the null hypotheses and that no further post-hoc testing is needed.

## Network Composition Comparison

We compared network composition by sector affiliation in terms of the alter categories of "Stage" and "Relationship," addressing RQ1B. Counts of alters classified in each category, split by affiliation with each sector, are shown in Tables 1 and 2 in the appendix. Since all networks were composed of alters that have some classification in both of these categories, the chi-squared test is appropriate for determining if any networks grouped by sector affiliation are different based on the relative frequency of the different classes of alter Stage and Relationship. Since chi-squared tests require mutually exclusive categorizations across the board, we ran three separate tests where each test compared a sector's IN group to its OUT group. In each case, results of these tests tell us whether we should reject the null hypothesis that there is no difference between a sector's IN and OUT groups in terms of alter composition.

For the Stage category, we found significant differences between the IN and OUT groups of the same two sectors, with both adjusted p-values $< 0.0001$. For the Relationship category, we found significant differences between the IN and OUT groups of Academia and Industry, with adjusted p-values $< 0.0001$ and $= 0.0020$, respectively. In order to determine exactly which factors were different, we ran 26 post-hoc tests for each alter category in Academia and Industry, which revealed the following significant differences for the Stage category:

- Participants in Academia had fewer grad**** alters and more none**** alters than those out of Academia.
- Participants in Industry had more grad**** alters than those out of Industry.

For alter composition based on Relationship:

- Participants in Academia had fewer friend*** alters and more peer** alters than those out of Academia.
- Participants in Industry had more friend* alters than those out of Industry.

More discussion of these differences will follow, but we visualize these different distributions here in Figures 6 and 7.

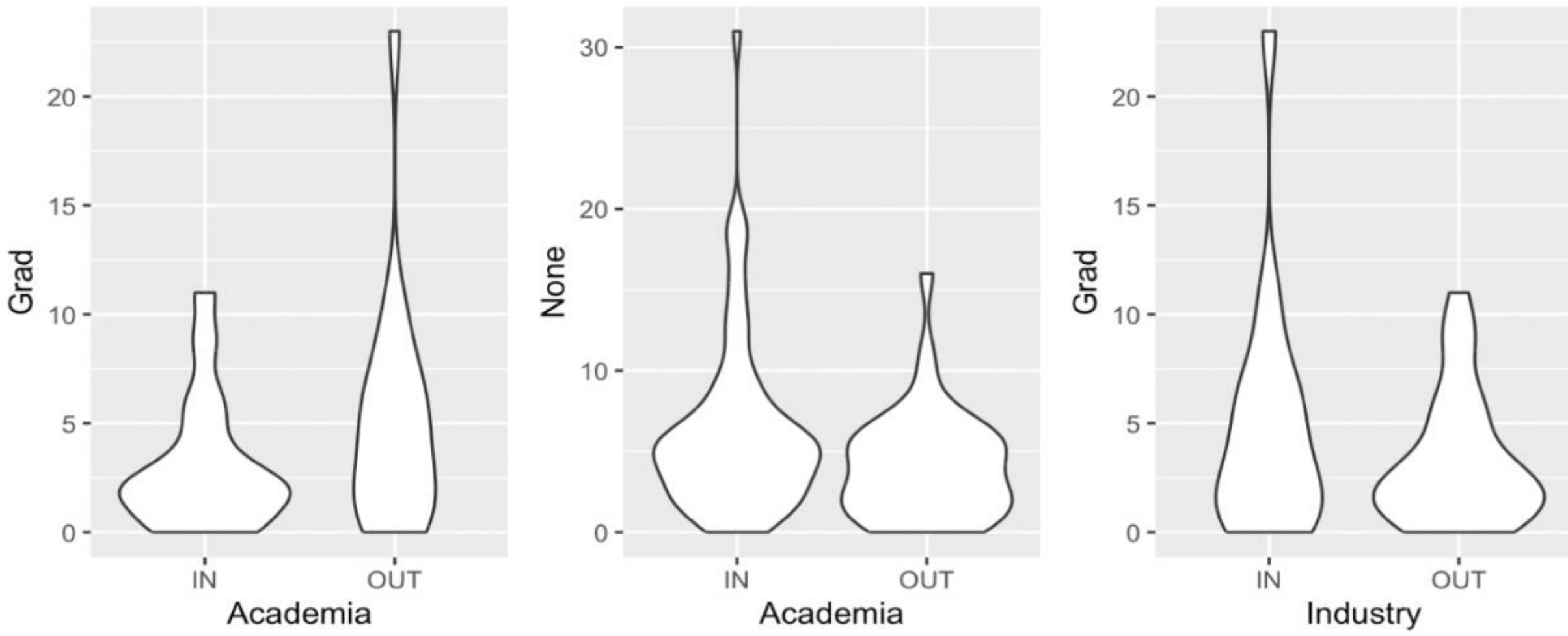

Figure 6: Violin plots comparing distributions of alter counts for groups found to be different based on p-values for the Stage category. Violins are normalized to have equal areas.

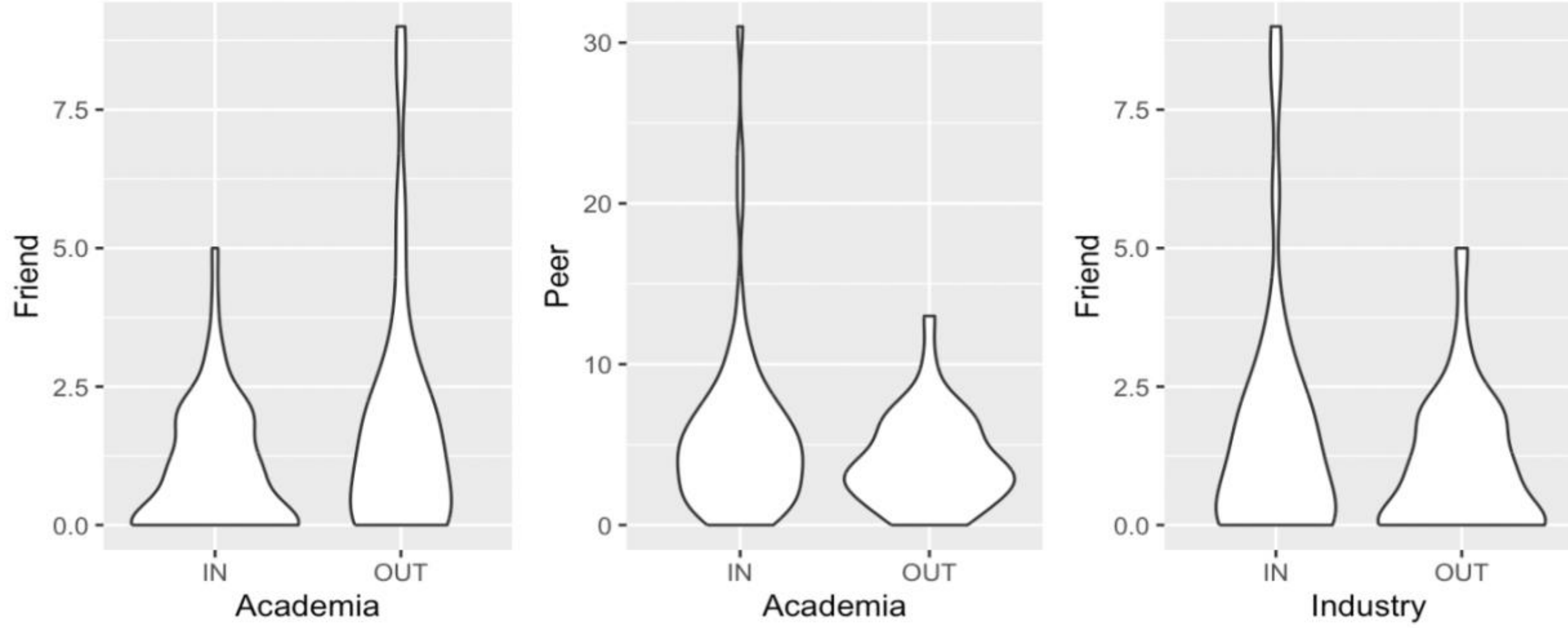

Figure 7: Violin plots comparing distributions of alter counts for groups found to be different based on p-values for the Relationship category. Violins are normalized to have equal areas.

## Support Type Comparison

We compared counts of different support types based on sector affiliation, addressing RQ1B. These counts are shown in Table 5 in the appendix. The chi-squared test of independence is appropriate for comparing each sector's IN and OUT groups on the basis of counts of associated support types. In this case, results of the tests tell us whether we should reject the null hypotheses, that each sector's IN and OUT groups are not different from each other based on relative frequencies of support types.

We found significant differences in terms of support type frequencies between the IN and OUT groups of all three sectors, with adjusted p-values < 0.0001 for Academia, = 0.00033 for Industry, and = 0.0079 for Government. We ran 27 post-hoc tests for each support type in all sectors, which revealed the following significant differences:

- Participants in Academia described less material support* than those out of academia.
- Participants in Industry described more networking support* than those out of Industry.
- Participants in Government described less instrumental support** than those out of Government.

We note that these conclusions apply in slightly different contexts. Overall, participants did not describe very much material support (less than half described any material support, let alone material support from multiple alters), so the first bullet should be read as those in Academia typically describe *no* material support while those out of Academia were more likely to describe *some* material support. On the other hand, all but two participants described some instrumental support (with most describing multiple sources of instrumental support), so the last bullet should be read as those in Government describing less, but still *some* instrumental support, while those out of Government described more instrumental support. Networking support, or the middle bullet, should be read as somewhere between these two interpretations. More discussion on these differences will follow, but we visualize these different distributions here in Figure 8.

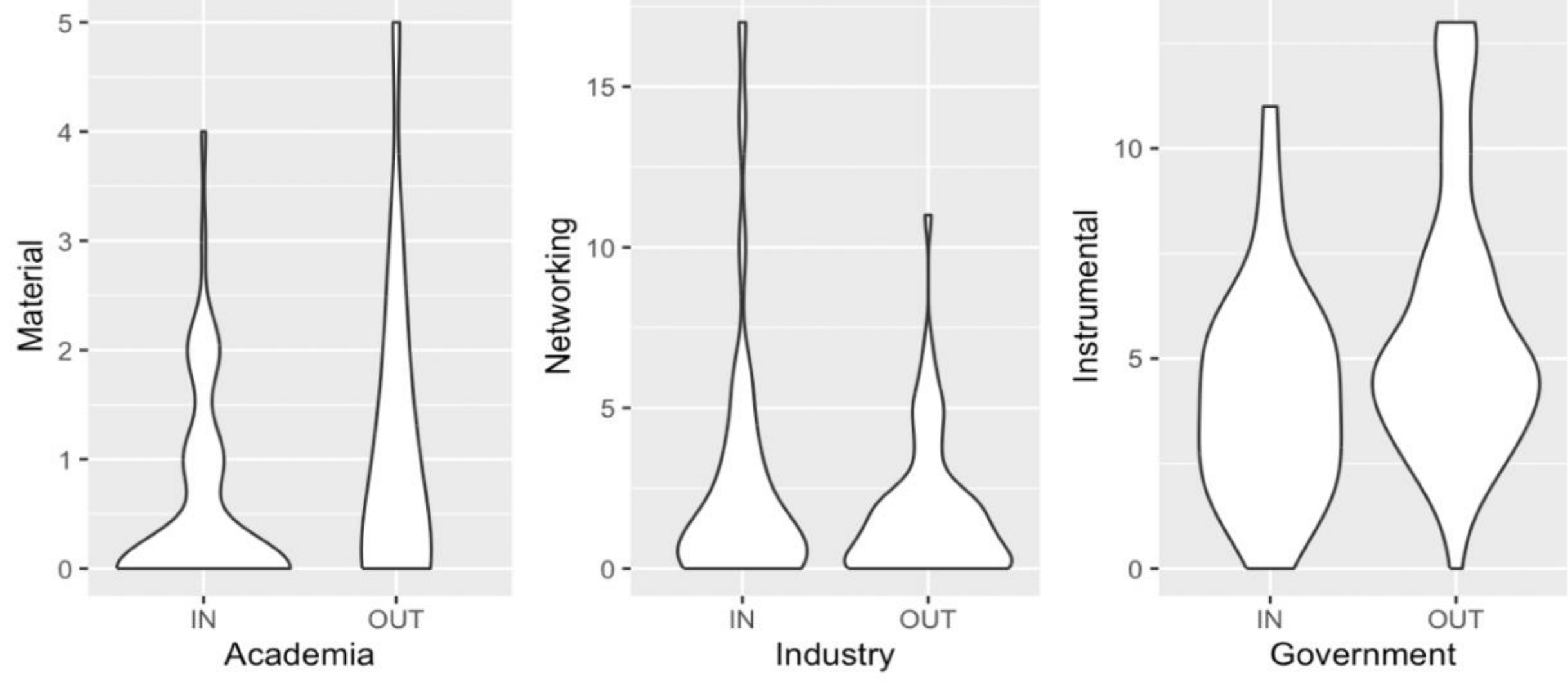

Figure 8: Violin plots comparing distributions of counts of support types for groups found to be different based on results of post-hoc testing. Violins are normalized to have equal areas.

# V. Discussion

## Interpretation and Implications of Full Group Results

Before discussing the implications of these results, we note again that the experiences of our participants may not be directly generalizable to the entire population of GSM physicists since our sample was not selected completely randomly. Furthermore, apart from comparisons from our sample made on the basis of career sector, our results do not allow us to make causal or correlational claims. However, we are able to comment on trends in our dataset, which may still reveal something about the experiences of GSM physicists more broadly, and we are able to connect observed trends to other research, giving clearer implications for our results.

### Networks and Metrics

Many aspects of participants' networks varied widely, which is apparent from the relative sizes of many standard deviation measures in Table 4. On average, each network had about 17 alters, but the number of alters ranged roughly normally from 4 to 47. Participants reported about 1.5 times as many instances of different types of support associated with different alters as the number of alters in their network, while they reported about one-tenth as many gaps in support as alters. Support measures varied, with some participants reporting more than three times as many support instances as alters while one participant reported more gaps in support than there were alters in their network. Interview questions were largely based on support (rather than gaps in support), so the abundance of support compared to gaps may not be surprising, but it is still worth noting that, overall, our participants were able to describe generally supportive networks.

Participants' networks were generally not very dense—about one fifth of possible connections existed in their networks, on average. This may not be surprising since the only thing necessarily linking alters in any network was their connection to the ego, and the ego is removed for meaningful density calculations. A tendency towards tight clustering when alter-alter connections are present is reflected by an average transitivity (also known as clustering coefficient) of 0.677, which suggests that groups of 3 alters featuring at least two connections tend to have all 3 connections more often than not.

Transitivity had an unexpected non-normal distribution (see Figure 9) that suggested a large tendency towards clustering, which we believe may be partially explained by sociogram construction instructions. After participants placed alters on sociograms, they were asked to indicate which alters knew each other by drawing lines connecting them or circling groups of alters who knew each other. When alters were circled, we counted each person in the circle as being connected to every other person in the circle. With this instruction usually coming at the end of a long interview, in some cases, and especially when sociograms contained many alters, participants may have opted for circling as an easier alternative to drawing each and every connection even if not all circled alters certainly knew each other. This is an example of a limitation of egocentric SNA described in the Methods Paper, where egos are an excellent source of information concerning their own connections but may be less reliable when it comes to their alters' connections. Regardless of the reasons for a tendency towards clustering, based on average density and transitivity values for the full group, it is apparent that participants' networks tended to feature connections between alters in groups of three or more, perhaps based on common institutional or collaborative affiliation.

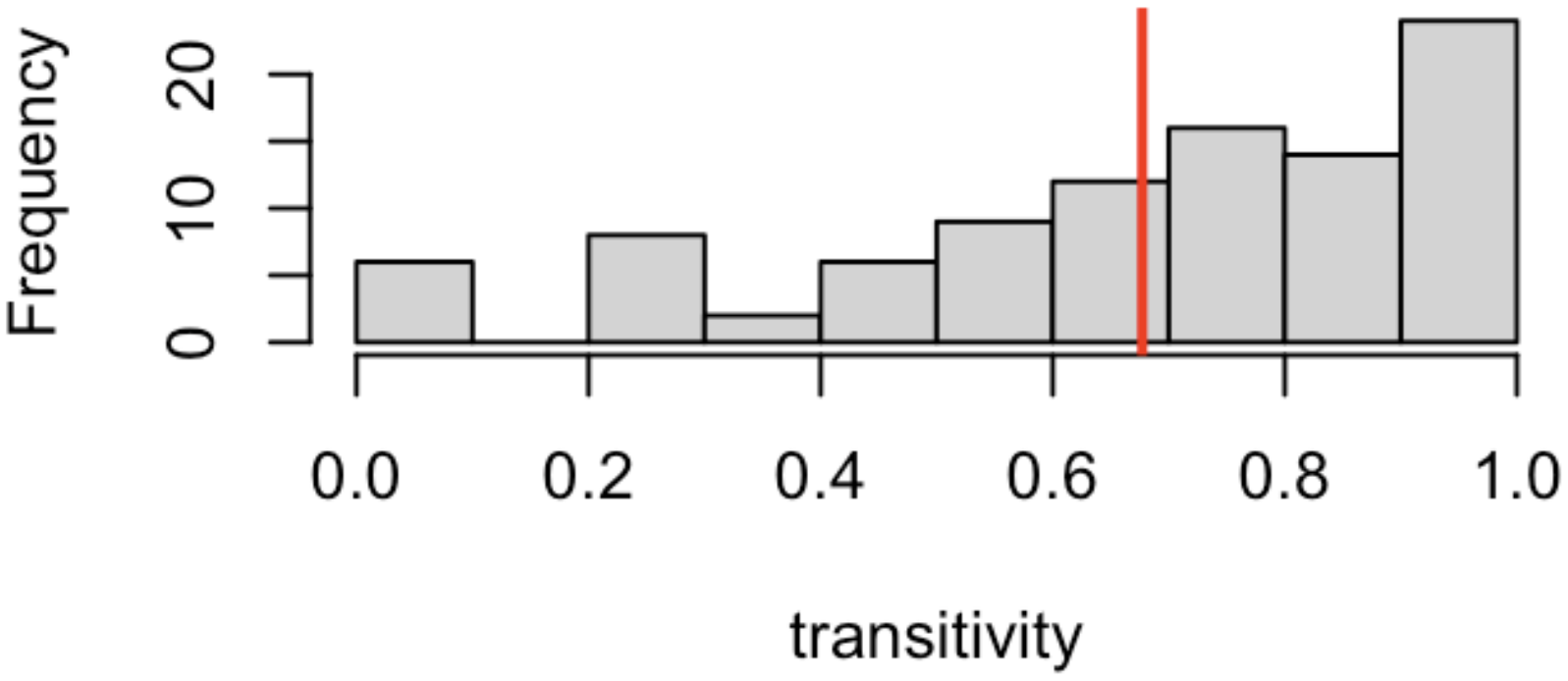

Figure 9: Histogram of transitivity values for the full group, showing non-normal distribution. The red line indicates the mean value.

Considering the networks of the full group, perhaps the most striking feature is the presence of support. With an average of 10 times as many instances of support recorded as gaps in support, we could say that our participants were able to describe more support in their professional journeys (compared with non-support), with emotional, instrumental, and career advice support most prevalent among them.

## Composition

In terms of career stage, participants' networks were mostly composed of alters they discussed first in the context of a job or who were not associated with a career stage ("none" category), about a third each. Many "none" category alters were personal friends, family, partners, or otherwise not formally associated with their ego's academic or professional careers. Participants first discussed many alters in the context of graduate school (approximately a quarter), with fewer alters first associated with their undergraduate or postdoc stages.

In terms of general relationship type, participants' networks were mostly composed of peers (over one quarter), bosses, and groups (less than a fifth and a sixth, each, respectively). The remainder included roughly equal representations of friends, professional friends, mentors, and family/partners at a little less than 10% each, though mentees were less common at about 4%.

Considering the composition of these physicists' networks, we see that connections made during jobs, graduate education, and outside of academic and professional institutions are critical to the extent of their professional networks. Likewise, peers, bosses and mentors, and connections associated with groups are foundational in their networks.

### Support

Each alter could be associated with 0 - 9 types of support. For the full group, we found that emotional, instrumental, and career advice support were most common at about a fifth of total support counts each. This may not be surprising since the interview protocol included questions which asked somewhat directly about these types of support. Material support and physical closeness were less common, representing less than 3% each of total support counts.

Bosses and connections made during graduate school were critical sources of networking support, meaning that those connections may have helped participants' networks grow and support them to the extent observed. Bosses, peers, and connections made at jobs were critical sources of instrumental support. Family, partners, friends (both personal and professional), and connections made through groups were critical to providing emotional support, meaning that most emotional support comes from outside the professional sphere or from choice professional connections, which is echoed in other research [36]. Connections made through groups, which included identity-based affinity groups, personal and professional interest groups, and more, were uniquely associated with providing identity-based and community building support as well as support through DEI initiatives and policies. This may not be surprising, since these types of support are often among desired outcomes of such groups, but it confirms that these groups generally accomplish these goals for our sample of GSM physicists.

The fact that group connections were the only connections providing a statistically significant share of these kinds of support also speaks to the unique and critical position of such groups in providing these kinds of support in general. The elimination or diminishment of DEI-related programs, which describes many of the groups discussed by our participants, reduces support for GSM physicists and raises the risk of their exit from the field, since these groups were unique in providing identity-based and community building support and since other studies have also identified these groups as critical to minority physicists' success [2,3,5,37].

## Interpretation and Implications of Sector Comparison Results

Analysis of our dataset has allowed us to draw some conclusions about differences in participants' experiences in the three career sectors. As discussed earlier, these conclusions may extend beyond our sample or not. We discuss them here as a way for (aspiring) professional GSM physicists to have a better idea of what to expect in the different sectors and as a way for employers in these sectors to consider what they may or may not be providing to their employees.

Comparisons between those in and out of Academia (jobs include faculty, research, administrative, and other roles at universities) suggest the prevalence of relationships with peers, in most cases at a job, for those in academia. Personal friends and connections made in graduate school do not feature as heavily in the support networks of those in academia, and material support is lacking. This last point may support a notion that the work itself (rather than the pay) is the reward in academia, a sentiment echoed in multiple interviews. Other points may suggest that professional life may take more precedence over personal life in academia compared to other sectors, a feeling also commonly alluded to by participants in academia. We found the relative lack of graduate school connections surprising, since graduate school is often a critical time in the life of an academic, but believe it makes more sense if we consider the way that the networks of those in Academia seem to prioritize immediate working and professional connections.

For those in Industry (various jobs in engineering, tech, science communication, consulting, and other fields), results suggest that personal friends and connections made in graduate school are more prevalent in their networks. They also see more networking support than others. Unlike academia, then, industry may be seen as a sector that is conducive to more investment in personal life and long-lasting connections, with a few participants in industry noting that they are not as prone to taking their work home with them. The relative abundance of networking support may be read as something that industry jobs provide, or it may be that those who received more networking support earlier on were better able to pivot out of academia and into industry after obtaining a PhD. We found the relative abundance of graduate school connections in Industry surprising, but believe it makes more sense when noted together with the relative abundance of networking support, some of which may have come from graduate school connections.

Participants in Government (positions at national labs, federal agencies, government-affiliated nonprofits, and others) describe less instrumental support, suggesting that those in the sector may find that they fend for themselves in terms of their work more than those in other sectors.

## Suggestions

The implications of results discussed above have their own implications for institutions that employ and support professional physicists. Generally, such institutions should recognize the ample and varying types of support described by these physicists and should consider how they can help connect their constituents with these types of support, especially instrumental, emotional, and career advice support. They should try to bolster workplace connections with supervisors and peers as effective means of receiving on-the-job support, perhaps through connection-building professional development or workplace social activities. They should recognize that emotional support, while not absent in professional environments, will largely come from the personal sphere and that constituents may need time and space away from work to access that sphere.

For institutions supporting and employing physicists across the three sectors, the importance of groups (like personal and professional interest groups, affinity groups, and identity-based groups) and the connections made therein can not be overstated. Our results show that, not only do these groups successfully meet their (assumed but likely) goals of providing identity-based and community building support to this sample of physicists, but they are the only entities and connections in our sample notable for these specific types of support. Depending on the groups, allocation of funding, time, space, and visibility may all be critical to the maintenance of these crucial support systems. Institutions should consider interest and affinity groups, whether professional or personal, as a necessary component in supporting GSM physicists. If considering eliminating or diminishing groups on the basis of DEI-relatedness, institutions should recognize that this would likely come at the expense of GSM physicists.

For sector-specific recommendations, we consider the implications of comparing participants' experiences in the three sectors. Employers in academia should continue to support connections within and across the institution (peers), but they should also consider the effect of a lack of material support and of a professional life that may leave less room for personal relationships. If more material support (which would include pay, benefits, access to funding, or other necessary resources) for employees is not an option, academic employers should consider changing expectations around time spent at work or otherwise continue to support workplace flexibility, also found by Whitten et al. to be key in the retention of women physicists [9]. Given other research that has identified support from outside professional/academic institutions and support from personal connections like family and spouses as critical to GSM physicists' success [5,15,36], it may be especially worthwhile for academic institutions to reframe "personal time" as a necessary dimension of professional success. Employers in industry should continue to support networking for their employees and should consider that work in their sector may be difficult to access for PhD physicists. They should continue to allow space for personal relationships to flourish, and should consider providing more opportunities for mentorship if desired by employees. Employers in government should consider providing more support and perhaps mentorship that could help their employees with their day-to-day work, which may mean more training, detailed instructions, or bolstering connections with bosses and peers.

## Limitations

In addition to limitations in the methods discussed in the Methods Paper (which included tradeoffs between qualitative descriptiveness and quantitative comparability, ambiguity in interview recordings, and variations in sociogram construction), this study is limited by a sample population that was not chosen completely randomly, which means we are unable to ensure generalizability to the larger population of GSM physicists. Still, by casting a wide net with invitations to join the study extended through national listservs and professional societies as well as social media advertisements, and with a relatively large sample, our results may at least be powerfully suggestive for the larger population of GSM physicists in the United States.

In some cases, we were limited by small population subsets when running statistical tests. Our need to base sector comparisons on sector affiliation rather than specific sector (combination) is an example of this. Consequently, our results are informative strictly on the basis of affiliation with the different career sectors and not necessarily on any participant's unique employment sector (combination). As with any meeting of qualitative description with statistical tools, there is a tradeoff between detail and nuance and explanatory power, and in this case we opted for explanatory power. Similarly, we note that the relatively small size of some groups compared to others (Undergrad (N=63) compared to Job (N=614) in the Stage category, for example) means we have less power to detect differences in those groups, especially considering our use of a conservative method to address multiple testing effects. Again, this is a trade off between finding fewer, more significant results and finding more, less significant results, and we have opted for the former.

Finally, this study is limited throughout, though usually minimally, by the perspectives, biases, and subjectivities of the various researchers that have interacted with the participants and data. As described in more detail in the Methods Paper, this is an unavoidable reality for qualitative data collection and analysis, though we have taken steps to mitigate it. Still, minor differences in interview techniques, thematic coding practices, and decisions regarding data classification have shaped our dataset in ways both predictable and not. As with any qualitative study, we do our best to account for these differences and we expect that they add some fuzziness, but not skew, to our results.

### Future work

For other researchers, we see an opportunity to undertake similar studies of the experiences of marginalized professional physicists, but on the basis of identities other than gender and sexual identities, like perhaps race, nationality, and (dis)ability status, as we found differences in experiences along these lines as well in this dataset.

We also see opportunities to look more closely at the differences between career sectors revealed by this work, specifically the differences in personal/professional relationships in networks and the prevalence of different kinds of support. It would be interesting to know if such differences exist for all physicists working in these sectors or just the ones in our sample.

Much more work from this project is forthcoming, and much of it is qualitative. On the quantitative side, we plan to examine different groups in our dataset with similar comparison schemes, like LGBTQ+ participants, Participants of Color, and participants with disabilities. On the qualitative side, our team will look at professional physics identity, the experiences of international participants, the experiences of participants with disabilities, and participants' understanding of closeness, among other topics.

## VI. Conclusion

We have presented the results of an extensive mixed-methods egocentric network analysis of the support networks of women and/or LGBTQ+ (gender and sexual minority, or GSM) PhD physicists, the methods for which are described in an earlier publication. We have reported descriptive results for the full group and comparative results for participants in or out of each of the three career sectors of academia, industry, and government. We have interpreted those results in light of what they suggest for our sample population and have made suggestions for institutions supporting and employing physicists.

This study and its results may serve as a guide for aspiring professional physicists, especially GSM physicists, on what they might expect when working in any of these sectors. It also may serve as a guide on the qualities of the networks of successful GSM physicists. This study also provides guidance and suggestions for institutions and organizations that support and employ professional physicists, particularly with regard to what types of support and what types of connections are important for professional GSM physicists. Ultimately, this work can help marginalized physicists make better decisions about their continued work in physics and can help physics institutions better support them.

# Acknowledgements

We would like to thank the study participants for their contributions to this work. This work was supported by NSF grants DUE-2100024, 2055237, and 2054920 until that funding was eliminated by the Trump administration. The work continues nonetheless.

# Appendix

## Network Composition Comparison

Table 1: Counts of alters in different classes of the Stage category, split by affiliation with each sector.

| **Sector (N)** | **Undergrad** | **Grad** | **Postdoc** | **Job** | **None** |
|---|---|---|---|---|---|
| $A_{IN}$ (1074) | 35 | 192 | 46 | 406 | 395 |
| $A_{OUT}$ (606) | 28 | 189 | 31 | 208 | 150 |
| $I_{IN}$ (565) | 16 | 170 | 33 | 184 | 162 |
| $I_{OUT}$ (1115) | 47 | 211 | 44 | 430 | 383 |
| $G_{IN}$ (535) | 26 | 132 | 25 | 199 | 153 |
| $G_{OUT}$ (1145) | 37 | 249 | 52 | 415 | 392 |

Table 2: Counts of alters in different classes of the Relationship category, split by affiliation with each sector.

| **Sector (N)** | **Boss** | **Fam.part** | **Friend** | **Friend.prf** | **Groups** | **Mentee** | **Mentor** | **Peer** |
|---|---|---|---|---|---|---|---|---|
| $A_{IN}$ (1074) | 177 | 80 | 59 | 109 | 143 | 49 | 105 | 352 |
| $A_{OUT}$ (606) | 117 | 51 | 66 | 57 | 87 | 12 | 68 | 148 |
| $I_{IN}$ (565) | 119 | 45 | 58 | 54 | 75 | 10 | 51 | 153 |
| $I_{OUT}$ (1115) | 175 | 86 | 67 | 112 | 155 | 51 | 122 | 347 |

| $G_{IN}$ (535) | 100 | 44 | 38 | 52 | 78 | 10 | 60 | 153 |
|---|---|---|---|---|---|---|---|---|
| $G_{OUT}$ (1145) | 194 | 87 | 87 | 114 | 152 | 51 | 113 | 347 |

## Support Type Comparison

Table 5: Counts of different types of support, split by affiliation with each sector.

| **Support type (N)** | **$A_{IN}$ (1074)** | **$A_{OUT}$ (606)** | **$I_{IN}$ (565)** | **$I_{OUT}$ (1115)** | **$G_{IN}$ (535)** | **$G_{OUT}$ (1145)** |
|---|---|---|---|---|---|---|
| ID (264) | 149 | 115 | 105 | 159 | 91 | 173 |
| Phys. Cl. (64) | 41 | 23 | 20 | 44 | 23 | 41 |
| Netw. (202) | 112 | 90 | 90 | 112 | 76 | 126 |
| Comm. (241) | 170 | 71 | 66 | 175 | 75 | 166 |
| Car. adv. (448) | 295 | 153 | 144 | 304 | 140 | 308 |
| Emo. (533) | 342 | 191 | 172 | 361 | 171 | 362 |
| Mat. (71) | 33 | 38 | 23 | 48 | 33 | 38 |
| Inst. (522) | 362 | 160 | 160 | 362 | 131 | 391 |
| DEI (145) | 97 | 48 | 59 | 86 | 52 | 93 |